\documentclass[aps, amsfonts, amsmath, nofootinbib,  byrevtex,
twocolumn, showpacs, floatfix
,prd]{revtex4}   
\usepackage{epsf}
\usepackage{graphicx}
\usepackage{bbm}
\usepackage{bm}
\usepackage{color} 
\def\uv{UV}

\newcommand{\rarr}{\rightarrow}

\newcommand{\no}{\noindent}
\newcommand{\be}{\begin{eqnarray}}
\newcommand{\ee}{\end{eqnarray}}
\newcommand{\hk}{\hspace{0.1cm}}

\newcommand{\rk}{\right)}
\newcommand{\lk}{\left(}

\newcommand{\fx}{\boldsymbol{x}}
\newcommand{\fq}{\boldsymbol{q}}
\newcommand{\fk}{\boldsymbol{k}}
\newcommand{\fl}{\boldsymbol{l}}
\newcommand{\fy}{\boldsymbol{y}}
\newcommand{\fA}{\boldsymbol{A}}
\newcommand{\fQ}{\boldsymbol{Q}}
\newcommand{\fB}{\boldsymbol{B}}
\newcommand{\fE}{\boldsymbol{E}}

\def\id{\mathbbm{1}}

\begin{document}

\title{ Subcritical solution of the Yang-Mills 
 Schr\"odinger equation in the Coulomb gauge} 
\author{D.~Epple, H.~Reinhardt, W.~Schleifenbaum}
\affiliation{Institut f\"ur Theoretische Physik\\
T\"ubingen University\\
Auf der Morgenstelle 14\\
D-72076 T\"ubingen\\
Germany}

\author{A.P.~Szczepaniak}
\affiliation{Physics Department and Nuclear Theory Center \\
Indiana University, Bloomington, IN 47405, USA. }

\date{\today}

\begin{abstract}
In the Hamiltonian approach to Coulomb gauge Yang-Mills theory, the
functional Schr\"odinger equation is solved variationally resulting in a set of
coupled Dyson-Schwinger equations. These equations are solved self-consistently
in the subcritical regime defined by infrared finite form factors. It is shown
that the Dyson-Schwinger equation for the Coulomb form factor fails to
have a solution in the critical regime where all form factors have
infrared divergent power laws. 
\end{abstract} 


\maketitle

\section{Introduction}
Low energy gluon modes are expected to be responsible for the  distinctive features of QCD such as confinement or chiral symmetry breaking. Current evidence for gluonic excitations 
 is sparse~\cite{Thompson:1997bs,Ivanov:2001rv,Szczepaniak:2003vg,Adams:1998ff}, however new experimental efforts at JLab, PANDA, and BES, through studies of hybrid and glueball spectra are expected to shed more light into the nature of gluonic excitations. It is thus highly desirable to develop a 
 theoretical framework for studies of the Yang-Mills sector that is both: rooted in QCD and practical. 
Coulomb gauge quantization  offers    such a
framework~\cite{Christ:1980ku,Swift:1988za,Schutte:1985sd,Cutkosky:1987yi,
Zwanziger:2003cf,Szczepaniak:2001rg,Szczepaniak:2003ve,Feuchter:2004mk,Reinhardt:2004mm}. 
Elimination of the longitudinal component of the gauge field by
imposing the Coulomb gauge condition, $\bm{\nabla} \cdot \fA^a({\fx})$
where $a$ denotes color ($a=1\cdots N_C^2 -1$), leaves only two physical
transverse components as independent  degrees of freedom. 
In the functional integral approach, the $A_0$ component becomes
constrained and defines the instantaneous potential of color charge,
while in the canonical quantization approach Weyl gauge, $A_0=0$, is
used and the static potential emerges from the resolution of the Gauss law.
   {In the Coulomb gauge  the resulting many-body system of transverse gluons interacting via long-range Coulomb exchange forces can be 
 studied using standard techniques. After ultraviolet (UV) regularization it is possible to introduce an ansatz  for the ground state vacuum wave functional in terms of the dynamical gluon variables and optimize it by varying the energy density.  In particular    with a  gaussian ansatz a canonical transformation exists which transforms the gauge fields to the particle representation.   These quasi-particles satisfy a dispersion relation with a mass gap that originates 
 from non-perturbative  self-interactions mediated by the long range
 Coulomb potential. The expectation value of the  Coulomb potential is
 computed self-consistently in the same vacuum state, and   these
 self-interactions lead to a strong enhancement of the potential for
 large separation between localized color charges. For a true linear potential the single quasi-gluon mass becomes infinite  which is  consistent with  what is expected for a confining phase:  formation of color states requires infinite energy. 
    In a  color singlet combination, however, residual interaction between constituents screens  
     the long range interactions and leads to a finite mass for bound states. A
     priori, however it is not guaranteed that a particular ansatz for the
     vacuum wave functional would result  not only in qualitative but also in
     quantitative description of all features of confinements.  The gaussian
     ansatz represents fluctuations around topologically trivial distributions of the gauge field and
      since topologically nontrivial configurations can also minimize
      the  non-abelian  Yang-Mills action, a more complicated vacuum
      wave functional may be needed to, e.g., reproduce the area law
      behavior of the spacial Wilson loop \cite{Greensite:1979yn,Greensite:2007ij}.  Also, the vacuum expectation value of the Coulomb potential   
      is actually not the same as the energy of the state with static sources.  
       External sources polarize vacuum and lead to string formation
       for large separations.  Thus one should not compare the
       expectation value of the Coulomb energy to the "Coulomb plus
       linear" potential from  lattice gauge simulations of temporal
       Wilson loops \cite{Greensite:2004ke,Greensite:2003xf}. 
 Nevertheless, the variational Coulomb gauge approach does reproduce the qualitative features of confinement  and studying  discrepancies between this approach and lattice results, can improve   
 our  understanding of the  underlying   many-body problem. 

In a series of papers we studied solutions of the variational problem for the 
vacuum and single quasi-particle properties under various 
approximations 
\cite{Szczepaniak:2001rg,Szczepaniak:2003ve,Feuchter:2004mk,Reinhardt:2007wh,
Reinhardt:2004mm,SchLedRei06}.
 The goal of this paper is to review these results, clarify the differences of
 the various approaches and resolve some open problems. Ref.\
 \cite{Szczepaniak:2001rg,Szczepaniak:2003ve} and
 \cite{Feuchter:2004mk,Reinhardt:2007wh} use different ans\"atze for the wave
 functions and differ in the extend to which the curvature in the space of gauge
 orbits introduced by Coulomb gauge fixing was included. While it was shown that
 the different ans\"atze for the variational wave functionals is irrelevant to
 the order of approximation \cite{Reinhardt:2004mm}, in particular for the
 infrared behaviour, the inclusion of the curvature of the space of gauge orbits
 is crucial for obtaining the correct infrared behaviour. The subject of the
 present paper is the following: We reconsider the renormalization of
 the Dyson-Schwinger (DS) equations
 resulting in the variational approach. In particular, we show that to the order
 considered all UV-divergencies can be removed by adding appropriate counter
 terms to the Hamiltonian.  While in previous calculations the horizon
 condition (i.e. an infrared diverging ghost form factor) was either assumed or at least  seem to follow from numerical solutions 
 in the
 present paper we study the coupled DS equations in the subcritical regime abandoning  
 the horizon condition. In particular,  we will use the infrared analysis of
 Ref.\ \cite{SchLedRei06} to investigate the criticality of the various
 DS equations, i.e.
 the disappearance of the solution of the DS equations as the infrared exponents exceed
 certain critical values. We will find that 
in the limit of critical coupling where the  ghost form factor becomes infrared singular,
the DS equation for the Coulomb form factor ceases to have a solution.
 This explains why no consistent solution for the Coulomb form factor was found
 when the horizon condition is implemented. 

The organization of the paper is as
 follows:
   In the following Section we summarize the approximations used in solving 
  Dyson-Schwinger equations of the  variational Coulomb gauge problem from
   earlier
   analyses. In Section ~\ref{Renormalization} we derive the DS equations
    and discuss various renormalization schemes. In Section~\ref{Infrared} we 
    discuss the infrared (IR) limit of the equations. Numerical results and 
    discussion is presented in Section~\ref{Numerical}. Conclusions and outlook 
    are given in Section~\ref{Summary}.

 \section{Variational Coulomb QCD}
 The set of Dyson-Schwinger (DS) equations for correlation functions 
 describing gluon, ghost and the Coulomb propagators are considered within a 
 variational approximation for the ground state  vacuum wave functional.  
 The Yang-Mills Coulomb gauge Hamiltonian $H=H(\fA,\bm{\Pi})$  is a function 
 of the transverse gluon field, $\fA^a(\fx)$ and the conjugated momenta, 
 $\bm{\Pi}^a(\fx)= - \fE^a(\fx)$, 
\begin{equation}
[\fA^a(\fx), \bm{\Pi}^a(\fy)] = i\delta^{ab}\delta_T(\fx-\fy)
\end{equation}
with ${\bf \delta}_T = (\id - \bm{\nabla}\bm{\nabla}/\bm{\nabla}^2)\delta$. 
In the Schr\"odinger representation, matrix elements of an operator 
${\cal O}(\fA, \bm{\Pi})$ are given by 
\begin{equation}
\langle \Psi' | {\cal O} |\Psi \rangle = \int {\cal D} A J(A)  \Psi'(A) 
{\cal O}\left[A, -i {{\delta} \over {\delta A}} \right] \Psi(A) \; .
\end{equation}
In ~\cite{Szczepaniak:2001rg,Szczepaniak:2003ve} the variational wave 
functional  $\Psi(A) = \langle A|0\rangle \equiv   \Psi_0(\omega_0,A)$  in form of a gaussian ansatz was 
used, 
\begin{eqnarray}
 \Psi_0(\omega_{0},A) &= &\exp\left(-{1\over 2} \int d\fx d\fy \fA^a(\fx) \omega_{0}(|\fx - \fy|) \fA^a(\fy)\right)  \nonumber \\
& =  & 
\exp\left(-{1\over 2} \int {{d\fk}\over {(2\pi)^3}} \fA^a(\fk) \omega_{0}(k) \fA^a(\fk)\right)  \nonumber \\
& \equiv &  \exp\left( -{1\over 2} \int A\omega_{0} A\right), \label{gauss} 
\end{eqnarray} 
where $\fA^a(\fk)$ is the Fourier transform of $\fA^a(\fx)$. The functional 
integration measure contains the Faddeev-Popov (FP) determinant, 
$J = \exp Tr \ln (-D\partial)$ where $D(\fx,a;\fy,b) = [\delta^{ab}\partial_{\fx}  
- g f_{abc} \fA^c(\fx) ] \delta(\fx-\fy)  $  is the covariant derivative in 
the adjoint representation. It reflects the curvature \cite{Feuchter:2004mk} of the Coulomb gauge 
field space.  In \cite{Feuchter:2004mk} it was proposed to include  this 
curvature in the definition of the  vacuum wave functional and use, 
  \begin{equation}
  \Psi(\omega_\alpha,A) = J^{-\alpha}(A) \Psi_0(\omega_\alpha,A) \label{wf} 
  \end{equation} 
  with $\alpha = \frac{1}{2}$, which is the usual ansatz for the ``radial'' wave
  function in curvilinear coordinates. In Ref.\ \cite{Reinhardt:2004mm}
it was shown that within the one loop approximation the resulting DS equations 
do not depend on $\alpha$. We will return to this point below. 
 The specific choice  $\alpha =1/2$  which has the 
    advantage  of eliminating $J$ from the  integration measure 
    when computing  vacuum expectation values (vev). 
  
   The DS equations involve expectation values of field operators. 
  The gluon two-point function 
   $(\bm{\delta}_T(\fk) = \id - \fk \fk/\fk^2)$ 
\begin{equation}
\int d\fx e^{i\fk\fx} \langle 0| \fA^a(\fx) \fA^b(0) |0\rangle = 
{{\delta^{ab}\bm{\delta}_T(\fk)}\over {2\omega(k)}}  
\end{equation} 
defines the gluon  gap function   $\omega (k)$, which also relates to the single gluon energy. 
  The instantaneous ghost propagator $d(k)$  (better say ghost two-point correlation function) is defined by 
  \begin{equation}
\delta^{ab}d(k)  = \int d\fx e^{i\fk\fx} \langle 0|{{\partial^2}\over {D\partial}}(\fx,a;0,b) |0\rangle. \label{ddef} 
\end{equation} 
The square of the FP operator $-1/(D\partial)$ enters the Coulomb potential.
 The Coulomb form factor  $f(k)$   measures the ratio of the { vev} of its square to the square 
  of vevs,
\begin{equation}
\delta^{ab} f(k) d^2(k)  =   \int d\fx e^{i\fk\fx} \langle 0| \left[  {{\partial^2 }\over {D\partial}} \right]^2(\fx,a;0,b) |0\rangle. 
\end{equation} 
Finally we also consider the expectation value of the curvature 
\cite{Feuchter:2004mk}
defined by 
\begin{equation}
\delta^{ab}\bm{\delta}_T(\fk) \chi(k) = -\frac{1}{2}\int d\fx e^{i\fk \fx} \langle 0 | 
{{ \delta^2 \ln J} \over {\delta \fA^a(\fx)  \delta \fA^b(0) }}
|0\rangle\; .
\end{equation}
The DS equations for these correlations functions were independently studied 
in \cite{Szczepaniak:2001rg,Szczepaniak:2003ve} 
and \cite{Feuchter:2004mk,Epple:2006hv} 
  under different approximation schemes. Below we summarize the derivation of 
  these equations 
  and comment on differences in their solutions obtained so far. More detailed
   analysis of the equations is presented in the sections that follow. 
  
  The gap equation for the gluon propagator is shown in Fig.\ ~\ref{gapfig} and 
  follows from minimizing the energy with respect to the   $\omega_\alpha$  in
  Eq.\ (\ref{wf}), 
\begin{equation}
0 = {{\delta} \over {\delta \omega_\alpha(k)}} \langle 0|H|0 \rangle \label{gap} 
\end{equation} 
In Fig.\ ~\ref{gapfig} the first two diagrams represent the contribution from the 
kinetic energy terms which include transverse gluon self-interactions from the  
chromo-magnetic energy proportional to $\fB^2$ and from chromo-electric energy 
which depends on the curvature and is proportional to $J^{-1} \fE J \fE$.   

\begin{figure}[h]
\includegraphics[width=2.5in]{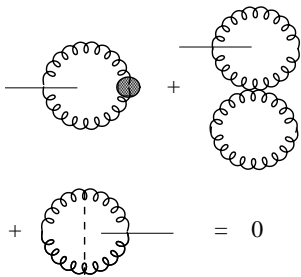}
\caption{ The gap equation.  The diagrams represent vev of the Coulomb gauge Hamiltonian. Thin solid line represent the derivative with respect to $\omega$. The blob represents expectation value of one-body operators, and the dashed line represents the expectation value of the Coulomb potential. \label{gapfig} } 
\end{figure}

 The exchange term represents the contribution from the Coulomb potential which 
 is determined by the expectation value of the Coulomb form factor $f$ and the 
 ghost propagator $d$. 
Given $\omega$ that solves the gap equation ~(\ref{gap}),  the ghost propagator is computed 
using Eq.\ (\ref{ddef}) by expanding the inverse of the FP operator in powers of 
the gluon field and summing all rainbow-ladder diagrams. These are the 
dominant contributions in the large $N_C$ limit. Corrections to the vertex, 
 coupling the transverse gluon and two ghosts that appears in the 
 expansion of  the inverse FP operator were estimated 
 in \cite{Szczepaniak:2001rg,SchLedRei06} and found to be small, of the 
 order
  of $O(10\%)$. We  postpone  discussion of renormalization to 
  Section \ref{Renormalization} but it is worth noting at this point that 
  in Coulomb gauge this vertex is UV finite. The diagrammatic representation 
  of the DS for the ghost propagator is shown in Fig.\ \ref{ghost}. 

\begin{figure}[h]
\includegraphics[width=2.5in]{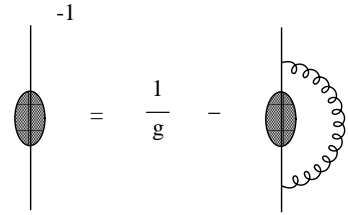}
\caption{ The DS equation for the ghost propagator, $d(k)$ represented by the oval. The gluon line is  represented by the $1/\omega$ propagator.\label{ghost} } 
\end{figure}
  
  The Coulomb form factor is treated in the same way and the resulting DS 
  equation is shown in Fig.\ \ref{coulomb}. Finally, to the same
  one-loop order the curvature is given by the loop shown in Fig.\
  \ref{chi} that is determined by the product of two ghost propagators
  arising from the expansion of $\delta^2 \ln J/\delta A \delta A$. 

\begin{figure}[h]
\includegraphics[width=2.5in]{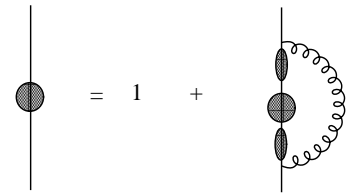}
\caption{ The DS equation for the Coulomb form factor $f(k)$ represented by the solid circle. The ghost propagator represented by the solid 
oval and the gluon line is represented by the $1/\omega$ propagator.\label{coulomb} } 
\end{figure}

\begin{figure}[h]
\includegraphics[width=2.5in]{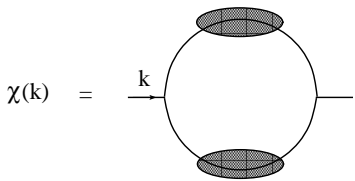}
\caption{ The one-loop contribution to the FP curvature, $\chi(k)$.  \label{chi}} 
\end{figure}

The resulting DS equations have a general structure,
\begin{equation}
F[k,f_i(k)] = \int d\fq K[f_j(\fk-\fq),\fk,\fq] \label{DSgen} 
\end{equation}
with $f_i$ being one of the four two-point function ($\omega$,$d$,$f$, $\chi$). 
While $\chi$ is merely a definition, $\omega$, $d$ and $f$ obey three
coupled DS equations. The main interest 
 is in establishing if these three equations have a common solution and if so what is 
 the  momentum dependence of the correlators, especially in the infrared limit.
   The bare DS equations contain UV divergences and have to be renormalized. 
   Since the variational method does not take into account 
  all diagrams contributing at a given order in the loop expansion removal of 
  ultraviolet divergences requires more than just dimensional transmutation of 
  the coupling constant. Renormalization of the coupling is  sufficient  to 
  render the ghost propagator finite and thus identifies the ghost propagator 
      with the running coupling. Renormalization of the Coulomb form factor 
      can be accomplished by resealing the Coulomb potential by a 
      renormalization scale dependent constant. The ghost loop is linearly 
      divergent and it turns out that  subtraction with a finite residual 
      scale dependence is necessary to render the gap equation  finite. 
 Thus renormalization of the curvature introduces a scale in addition to the 
 one defined by the renormalized coupling. As discussed earlier additive change of the 
 curvature  is equivalent to  a change in $\omega_\alpha$. In particular the 
 inverse gluon propagator is given by \cite{Reinhardt:2004mm}
  \begin{equation}
  \omega(k) =  \omega_\alpha(k) + (1 - 2\alpha) \chi  \label{ach}
  \end{equation}
 For our preferred choice $\alpha=1/2$ the wave functional parameter
 $\omega_{1/2}=\omega(k)$ becomes 
 identical with the inverse of the gluon propagator.  It follows from 
 Eq.\ (\ref{ach}) that a scale-dependent counter-term  in $\chi$ 
  can be absorbed into  $\omega_\alpha$ and thus can be interpreted as part 
  of the definition of the variational wave functional ansatz. We discuss this in 
  more detail in the following section. Finally there are quadratic and linear
     divergences  in the gap equations which can be removed by the  
     dimension-2 and dimension-3 counter-terms proportional to $\fA^2$ and 
     $\{ \fA, \bm{\Pi}\}$, respectively. To summarize, the variational 
     approach introduced three parameters. One is the scale that can be 
     associated with $\omega_\alpha$ as part  of the wave functional definition 
     or alternatively it can be associated with $\chi$ where it represents 
     finite renormalization. The second is the value of the coupling constant
   at some arbitrary value of momentum $d(k=\mu_d)$, and third is  the value 
   of the Coulomb form factor at an arbitrary value of momentum $f(k=\mu_f)$. 
  
The first study of the DS equations in this framework was carried out in 
Ref.\ \cite{Szczepaniak:2001rg} under the approximation $\chi = 0$, $\alpha=0$,
 so that $\omega(k) = \omega_0(k)$. 
   The remaining three equations were solved analytically under angular 
   approximation,
\begin{equation}
|\fk - \fq| \to k \theta(k - q) + q \theta(q - k) 
\end{equation}
for the integral on the r.h.s.\ of Eq.\ (\ref{DSgen}). Furthermore, the 
solution of the gap equation was approximated by $\omega(k) = 
m_g \theta(m_g- k) + k\theta(k - m_g)$. The effective gluon mass $m_g$  
plays the role of the scale parameter discussed above. The solutions for 
$d(k)$ and $f(k)$ were found to 
 exist for  the renormalized coupling $g(\mu)\equiv d(k=\mu)$ less than a 
 critical value. If the renormalization scale is  chosen  to be $m_g$ then 
 $g(m_g) <  4\pi\sqrt{9/(40N_C)}$. For sub-critical couplings both $f(k)$ 
 and $d(k)$ are finite in the limit $k\to 0$. At the critical coupling $d(k)$ 
 and $f(k)$  diverge as $1/\sqrt{k/m_g}$.  Similar behavior of the  numerical
  solutions without angular approximations was also found. For the numerical 
  solutions,  however, it could not be verified if in the limit $g \to g_c$ 
  $d(k)$ and $f(k)$ were also IR divergent. This was assumed to be given and 
  accordingly an analytical approximation was proposed.  
  In \cite{Szczepaniak:2003ve} the effect of the curvature was partially 
  included and a 
  numerical  sub-critical solution to all three equations was found (albeit 
  under some approximations to the gap equations). No attempt was made to 
  establish the IR limit nor the value of the critical coupling.  At this 
  point it is worth mentioning that the existence of the critical coupling may be 
  an artifact of the variational approximations and the rainbow-ladder  
  approximation.  Functional integration over the Coulomb gauge fields ought 
  to be restricted to the fundamental modular region where $Det(-D\partial)$ 
  is positive. In our approximation,  however, in each diagram gauge field 
  integration is 
  unrestricted. Thus it  is indeed expected that the sum over an infinite set of  
  diagrams, (e.g.\ of the rainbow-ladder series) converges only in a 
  restricted range of the coupling constant. 
  
The IR behavior of the Coulomb gauge correlators beyond angular approximation
and with full inclusion of the curvature
 was for the first time extensively studied in \cite{Feuchter:2004mk}.  There 
 it was found that in the IR the gap equation  forces the gluon propagator
   to have the same momentum behavior as the curvature, $\omega(k) 
   \sim \chi(k)$. Furthermore, under the approximation $f(k)=1$ it was shown 
   that the gap equation and the ghost DS equation admit a solution with 
   $\omega(k) \to \infty$ as $k \to 0$. Such a solution was not found
 previously in \cite{Szczepaniak:2001rg,Szczepaniak:2003ve}. In summary, the 
 full set of three DS equations have not been solved so far. The differences in
  the various solutions currently available may be related to the $f(k) = 1$ 
  approximation (in the  IR  limit) in \cite{Feuchter:2004mk}  or the 
  approximation $\chi=0$ used  in \cite{Szczepaniak:2001rg}, or  other 
  approximations made in \cite{Szczepaniak:2003ve}. 
 
 In the following we analyze the full set of equations, explain why 
 in ~\cite{Feuchter:2004mk} it was possible to find an IR vanishing gluon 
 propagator  while it was not the case for the solution found 
  in ~\cite{Szczepaniak:2001rg,Szczepaniak:2003ve}, and present a full  
  set of numerical solutions to all three equations. 

\section{Renormalization in the  Hamiltonian approach}
\label{Renormalization} 
Each one of the Dyson-Schwinger equations, for the ghost propagator, $d$, the 
Coulomb form factor $f$,  the gap equation for $\omega$ and the curvature 
$\chi$ requires subtraction of the UV divergences. 
The gap equation and the Faddeev-Popov determinant contain power divergences 
and we concentrate on those first. 

\subsection{Renormalization of the Faddeev-Popov determinant}
Within the approximation we are working the Faddeev-Popov determinant
\be
\label{9}
J (A) = Det (- \hat{D} \partial ) = \exp \lk Tr \ln (- \hat{D} \partial) \rk
\ee
can be replaced by \cite{Reinhardt:2004mm}
\be
\label{10}
J (A) = \exp (- \int A \chi A) \hk ,
\ee
i.e. from the Faddeev-Popov determinant mainly the curvature $\chi [\Psi] (k)$
enters. Note that the curvature depends on the chosen wave functional $\Psi$
and the representation Eq.\ (\ref{10}) can be used only inside the vacuum expectation
value (in the state $\Psi$) and to the considered order (two loops in the
energy). From the representation in Eq.\ (\ref{10}) it is clear that the
 counter terms
required to renormalize the Faddeev-Popov determinant, or more precisely its
logarithm, has to be of the form 
\be
\label{11}
\sim \int A A \hk .
\ee
Since the log of the Faddeev-Popov determinant can be considered as
part of the ``action'', we renormalize the Faddeev-Popov determinant as
\begin{eqnarray}
\label{12}
J \to J \Delta J & = & \exp \left[ Tr \ln (- D \partial) \right. \nonumber \\ 
&& \left.  - C_\chi (\Lambda) \int d\fx \lk \fA^a (\fx) \rk^2 \right]
\end{eqnarray}
or by using the representation Eq.\ (\ref{10}), we obtain
\be
\label{13}
J \Delta J = \exp \left[ - \int A \lk \chi - C_\chi (\Lambda) \rk A \right] 
\hk.
\ee
Obviously the counter term $C_\chi (\Lambda)$ has to be chosen to
eliminate the ultraviolet divergent part of the curvature $\chi (k)$. 
Thus the renormalization condition reads in momentum space
\be
\label{14}
\chi (k) - C_\chi (\Lambda) = \hk \mbox{finite} 
\hk .
\ee
As usual there is
 some freedom in choosing the finite constants of the right hand
side of Eq.\ (\ref{14}). In principle, we could just eliminate the ultraviolet
divergent part of $\chi (k)$. Since $\chi (k)$ has dimension of momentum and
furthermore $\chi (k)$ is linearly divergent in the ultraviolet, it is clear
that the ultraviolet divergent part of $\chi (k)$ is given by $\Lambda \cdot
const.$. So in principle, we could restrict ourselves to remove 
 just the \uv-divergent part of $\chi (k)$ by appropriately choosing 
 the counter
term $C_\chi (\Lambda)$. 
However,  it is more convenient to choose $C_\chi
(\Lambda)$ to be the curvature at some renormalization scale $\mu$, 
resulting in
the renormalization condition 
\be
\label{**}
C_\chi (\Lambda) = \chi (\mu)
\ee
 and in the finite
renormalized curvature
\be
\label{15}
\bar{\chi} (k) = \chi (k) - \chi (\mu) \hk .
\ee
It is easy to check that this quantity is indeed ultraviolet finite and
obviously it satisfies the condition
\be
\label{16}
\bar{\chi} (k = \mu) = 0 \hk .
\ee
By adopting the renormalization condition Eq.\ (\ref{**})
the renormalized quantity $\bar{\chi} (k)$ Eq.\ (\ref{15}) depends on the so far
arbitrary scale $\mu$. By choosing the renormalization condition Eq.\ (\ref{**}) this
renormalization scale becomes a parameter of our ``model'', since it defines the
infrared content of the curvature $\chi (k)$, which we keep in the
renormalization process, see Fig.\  \ref{g18} 

\begin{figure}[h]
\includegraphics{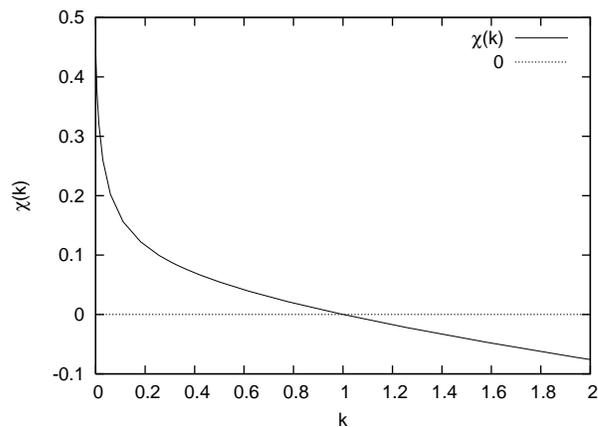}
\caption{The scalar curvature. Note the zero of $\bar\chi(k)$ at the renormalization scale $\mu=1$.}
\label{g18}
\end{figure}

If we choose for example $\mu = 0$, we chop off the whole infrared divergent
part of the curvature.

In Ref.\ \cite{Reinhardt:2007wh} we have chosen a different renormalization
condition keeping from the ultraviolet divergent quantity $\chi (\mu)$ the
finite part $\chi' (\mu)$. The renormalized curvature then reads
\be
\label{19}
\chi (k) = \bar{\chi} (k) + \chi' (\mu) \hk .
\ee
This implies that we subtract only the divergent part of the curvature keeping
fully its finite part. Then we have an extra parameter $\chi' (\mu)$ of the
theory and the renormalization scale $\mu$ is not related to the zero of the
renormalized quantity ${\chi} (k)$ Eq.\ (\ref{19}). Both points of view can be
adopted:

\begin{enumerate}
\item Using the renormalization condition of Eq.\ (\ref{**}) resulting in Eq.\ (\ref{16}) and
the renormalization scale $\mu$ becomes a physical parameter of the theory.
\item Removing only the \uv-divergent part of the curvature $\chi (\mu)$ keeping
its finite part $\chi' (\mu)$, which is equivalent to choosing the
renormalization condition
\be
\label{20}
C_\chi (\Lambda) = \chi (\mu) - \chi' (\mu)
 \hk ,
\ee
we obtain an extra finite renormalization  parameter $\chi' (\mu)$ and the renormalization scale $\mu$
remains arbitrary.
\end{enumerate}

\subsection{Counter terms to the Hamiltonian}
According to the general principles of renormalization we should introduce
counter terms to the Hamiltonian, calculate the expectation value of the energy
and then choose the coefficients of the counter terms such that all divergences
disappear from the gap equation. For the Yang-Mills Hamiltonian in Coulomb gauge
and the wave functional at hand it turns out the counter terms required are of
the form
\begin{eqnarray}
\label{1}
\Delta H & = & C_0 (\Lambda) \int d\fx \lk \fA^a (\fx) \rk^2 \nonumber \\
 & &  + C_1 (\Lambda) \int d\fx  \fA^a (\fx) \cdot \bm{\Pi}^a (\fx) \hk . 
\end{eqnarray}
Here the coefficients $C_i (\Lambda) , i = 0, 1$ depend on the momentum cutoff
$\Lambda$ and have to be adjusted so that the 
\uv-singularities in the gap equation
disappear. Note also that these coefficients multiply ultralocal operators which
are singular in quantum field theory.

We will assume the wave functional of Eq.\ (\ref{wf}) with $\alpha=1/2$, as in \cite{Feuchter:2004mk}. For this wave functional the expectation value of the counter term
Eq.\ (\ref{1})  is given by
\be
\label{2}
\Delta E & = & C_0 (\Lambda) \frac{1}{2} t_{ii} \delta^{aa} \int d^3 \fx \omega^{- 1} (\fx,
\fx) \nonumber \\
 &+ &  i C_1 (\Lambda) \int d^3 \fx \langle \fA^a (\fx) \cdot \tilde{\bm{\Pi}}^a (\fx)
\rangle_\omega \hk ,
\ee
 where $\langle \dots \rangle_\omega$ denotes the expectation value in the
 Gaussian wave functional with kernel $\omega=\omega_{1/2}$ and 
\be
\label{3}
\tilde{\bm{\Pi}}^a (\fx) = J  \raisebox{3mm}{\tiny 1} \hspace{ -1mm }
\raisebox{2mm}{\tiny /} \hspace{ -1mm }
\raisebox{1mm}{\tiny 2} \bm{\Pi}^a (\fx) J^{-} \raisebox{3mm}{\tiny 1} 
\hspace{ -1mm }
\raisebox{2mm}{\tiny /} \hspace{ -1mm }
\raisebox{1mm}{\tiny 2} = \bm{\Pi}^a (\fx) - \frac{1}{2} \lk \bm{\Pi}^a (\fx) \ln J \rk 
\hk .
\ee
In Ref.\ \cite{Feuchter:2004mk} it was shown
\begin{eqnarray}
\label{4}
& & \langle \fA^a (\fx) \cdot \tilde{\bm{\Pi}}^a (\fx) \rangle_\omega \equiv  i \langle
\fA^a (\fx) \cdot \fQ^a (\fx) \rangle_\omega \nonumber\\
& = & i \int d \fx_1 \omega^{- 1} (\fx, \fx_1) \lk \omega (\fx_1, \fx) - \chi (\fx_1, \fx)
\rk \delta^{a a}  \nonumber \\
\end{eqnarray}
where $\chi (k)$ is the curvature and $\delta^{aa} = N^2_C - 1$. In momentum
space we find then for the expectation value of the counter terms (\ref{2})
\begin{eqnarray}
\label{5}
\Delta E  &= & \lk N^2_C - 1 \rk V \left[ C_0 (\Lambda) \int \frac{d \fk}{(2 \pi)^3}
\frac{1}{\omega (k)} \right.  \nonumber \\
& & \left. - C_1 \int \frac{d\fk}{(2 \pi)^3} \frac{\omega (k) - \chi
(k)}{\omega (k)} \right] 
\hk .
\end{eqnarray} 
Taking the variation of this expression with respect to $\omega (k)$ we obtain
\begin{eqnarray}
\label{6}
\frac{\delta \Delta E}{\delta \omega (k)} & =  & \lk N^2_C - 1 \rk \frac{V}{(2
\pi)^3}  \nonumber \\
& & \left[ -C_0 (\Lambda) \frac{1}{\omega^2 (k)} - C_1 (\Lambda) \frac{\chi
(k)}{\omega^2 (k)} \right] \hk .
\end{eqnarray}
Note that $\chi (k)$ depends only on the ghost propagator but not on the gluon
 gap function   $\omega$, at least as long as $\chi (k)$ is not yet the self-consistent
solution.

\no Variation of the expectation value of the energy (without counter terms) yields
\begin{eqnarray}
\label{7}
\frac{\delta E}{\delta \omega (k)}  &= &  \frac{N^2_C - 1}{2} {\delta}^3 (0)
  \nonumber \\
& \times &\frac{1}{\omega^2 (k)} \left[ - k^2 + \omega^2 (k) - \chi^2 (k) - I^0_\omega -
I_\omega (k) \right], \nonumber \\
\end{eqnarray} 
where $\delta^3 (0) = \frac{V}{(2 \pi)^3}$ and all other quantities are defined
in Ref.\ \cite{Feuchter:2004mk}. Adding the counter terms from Eq.\ (\ref{5}) to the energy we obtain the gap equation 
\begin{eqnarray} 
\label{8}
\omega^2 (k) & - &  \chi^2 (k) =  \nonumber \\
 & = & k^2 + I^0_\omega + I_\omega (k) - 2 C_0 (\Lambda) -
2 C_1 (\Lambda) \chi (k). \nonumber \\
\end{eqnarray}

\subsection{Renormalization of the gap equation}
We now turn to the renormalization of the gap equation  (\ref{8}), which
includes already the counter terms. After the renormalization of the
Faddeev-Popov determinant, the curvature $\chi (k)$ can be replaced by the
renormalized one $\bar{\chi} (k)$, Eq.\ (\ref{15}). Note that the Coulomb integral
$I_\omega (k)$ 
does not depend on any constant part of the curvature, so that $\chi (k)$ could
have been replaced right away by the finite quantity $\bar{\chi} (k)$, 
Eq.\ (\ref{15}). Replacing
$\chi (k)$ by $\bar{\chi} (k)$ and using the relation, see Ref.\
\cite{Feuchter:2004mk}
\be
\label{21}
I_\omega (k) = I^{(2)}_\omega (k) + 2 \bar{\chi} (k) I^{(1)}_\omega (k)
 \hk ,
\ee
the gap equation in Eq.\ (\ref{8}) becomes 
\begin{eqnarray}
\label{22}
& & \omega^2 (k) -   \bar{\chi}^2 (k) =  \nonumber \\
& = & k^2 + I^0_\omega + I^{(2)}_\omega (k) - 2 C_0
(\Lambda) + 2 \bar{\chi} (k) \lk I^{(1)}_\omega - C_1 (\Lambda) \rk \hk \nonumber \\
\end{eqnarray}
The integrals $I^{(n = 1, 2)}_\omega (k)$ are linearly $(n = 1)$ and
quadratically $(n = 2)$ \uv-divergent and are defined in \cite{Feuchter:2004mk}. 
Furthermore, the integral $I^0_\omega$ is also quadratically
divergent but independent of the external momentum. We can therefore eliminate
all \uv-divergences by choosing $C_0 (\Lambda) \sim \Lambda^2$ and $C_1
(\Lambda) \sim \Lambda$. In principle, the coefficients of the counter terms, $C_i (\Lambda) , i = 0, 1$,
have to be chosen to eliminate the \uv-divergent parts of the quantities
appearing in the gap equation. This means that we should choose the infinite
coefficients $C_i (\Lambda) , i = 0. 1$ as
\be
\label{26}
\lk I^0_\omega + I^{(2)}_\omega (k) \rk_{\mbox{\uv-divergent part}} - 2 C_0
(\Lambda) & = & 0 \nonumber \\ 
&& \label{27}  \\
\left. I^{(1)}_\omega  (k) \right|_{\mbox{\uv-divergent part}} -  C_1
 (\Lambda) &
= & 0. \nonumber \\
\ee
Note that the \uv-divergent parts of $I^{(n = 1, 2)}_\omega (k)$ are by
dimensional arguments independent of the external momentum $k$. Technically it
is more convenient to choose the following alternative renormalization conditions.
 As usual we have the freedom in choosing the
renormalization conditions up to finite constants. Given the fact that
$I^0_\omega$ is independent of the external momentum and the differences 
\be
\label{23}
\Delta I^{(n)}_\omega (k, \nu) = I^{(n)}_\omega (k) - I^{(n)}_\omega (\nu)
\ee
are \uv-finite, we can eliminate all \uv-divergences by choosing the
renormalization conditions
\be
\label{24}
I^0_\omega + I^{(2)}_{\omega} (k = \nu) - 2 C_0 (\Lambda) &  = & 0  \\
\label{25}
I^{(1)}_\omega (k = \nu) - C (\Lambda) & = & 0 \hk ,
\ee
where $\nu$ is an arbitrary renormalization scale, which could be chosen to be
the same scale $\mu$ of the renormalization of the curvature, but given the fact
that with the renormalization presciption Eq.\ (\ref{**}), the scale $\mu$ becomes a
physical parameter, the two renormalization parameters $\nu$ and $\mu$ need not
necessarily  be the same. With the renormalization conditions Eqs.\ (\ref{24},\ref{25}) the renormalized
(finite!) gap equation reads
\be
\label{28}
\omega^2 (k) - \bar{\chi}^2 (k) = k^2 + \Delta I^{(2)}_\omega (k, \nu) + 2
\bar{\chi} (k) \Delta I^{(1)}_\omega (k , \nu) \hk .
\ee

Assuming that the integrals $I^{(n = 1, 2)}_\omega (k)$  are infrared finite 
and
furthermore that the renormalized curvature $\bar{\chi} (k)$
  is infrared divergent,
the infrared limit of the renormalized gap equation is given by
\be
\label{29}
\lim\limits_{k \to 0}
 \lk \omega (k) - \bar{\chi} (k) \rk = \Delta I^{(1)}_\omega (k = 0,
\nu) 
 \hk  .
\ee
To get the perimeter law of the 't Hooft loop
 requires that the quantity on the
right hand side of Eq.\ (\ref{29}) vanishes. This compells us to choose the, in
principle, free renormalization parameter $\nu$ as $\nu = 0$ \cite{Reinhardt:2007wh}. With this choice
the renormalized gap equation becomes
\begin{eqnarray} 
\label{30}
&& \omega^2 (k) - \bar{\chi}^2 (k) = k^2 + \Delta I^{(2)}_\omega (k, 0) + 2
\bar{\chi} (k) \Delta I^{(1)}_\omega (k, 0). \nonumber \\
\end{eqnarray}
The coupled DS equations then contain the following so far
undetermined parameters: the renormalization scale $\mu$ of the
renormalization of the curvature and the renormalization constants of the ghost
and Coulomb
form factors, the former one is fixed by the horizon condition. 

\subsection{ Renormalization of the ghost and Coulomb form factor} 
The Dyson-Schwinger equation for the ghost propagator, $d$ is given by 
 \cite{Feuchter:2004mk}
\begin{equation} 
{1\over {d(k)}} = {1\over g} - {N_C \over 2} \int d\fl {{ 1- (\hat\fl \cdot \hat \fk)^2 } \over {\omega(l)}} 
{ {d(\fl -\fk)} \over  {(\fk - \fl)^2}} 
\end{equation} 
and is renormalized by noticing that it corresponds to a running coupling. 
Regularizing the $\fl$ integral in the UV by a cutoff $\Lambda$ and replacing 
$g \to g(\Lambda)$ leads to a finite equation for $d$. It is convenient to 
renormalize $d$ at an IR rather than UV scale which can be done by 
subtraction, 

\begin{eqnarray}
\label{d} 
{1\over {d(k)}} = {1\over {d(\mu)} }  &- & \left[ {N_C \over 2} \int d\fl {{ 1- (\hat\fl \cdot \hat \fk)^2 } \over {\omega(l)}}  
{ {d(\fl -\fk)} \over  {(\fk - \fl)^2}} \right. \nonumber \\
& - & \left. ( k \to \mu) \right] \; .
\end{eqnarray}
The Dyson equation for the Coulomb form factor is given by 
\begin{equation} 
f(k)  = Z_f +  {N_C \over 2} \int d\fl {{ 1- (\hat\fl \cdot \hat \fk)^2 } \over {\omega(l)}} 
{ {d^2(\fl -\fk)f(\fl - \fk) } \over  {(\fk - \fl)^2}} \; .
\end{equation} 
Here $Z_f = Z_f(\Lambda)$ is a Coulomb kernel renormalization constant and the equation renormalized at an IR scale $\mu$ is given by 
\begin{eqnarray} 
\label{f} 
f(k)  = f(\mu) & + &   \left[ {N_C \over 2} \int d\fl {{ 1- (\hat\fl \cdot \hat \fk)^2 } \over {\omega(l)}} 
{ {d^2(\fl -\fk)f(\fl - \fk) } \over  {(\fk - \fl)^2}}  \right. \nonumber \\
& & \left.  - (k \to \mu) \right] \; .
\end{eqnarray}

\section{Infrared analysis at critical coupling}
\label{Infrared}
Here, it is shown that, within the approximations made, there exists
no simultaneous solution to the integral equations for $d$, $\omega$,
and $f$ at critical coupling for which the ghost form factor becomes infrared
divergent. We first present a simultaneous solution
of the integral equations for the ghost form factor $d(p)$ and the
Coulomb form factor $f(p)$ as defined above in  Eqs.(\ref{d}),~(\ref{f}). 
The result
obtained is not compatible with the solution obtained by simultaneous treatment
 of ghost and gluon DS equations found in \cite{SchLedRei06}. 
We want to check if there are solutions of the Dyson equations 
for  $\omega(p)$, $d(p)$ and $f(p)$ with the ghost form factor $d(p)$
being IR enhanced. To this end, we set for $p \to \infty $
\be
\label{ir-ansatz-schl-d}
\frac{1}{2}\omega^{-1}(p)=\frac{A}{(p^2)^{1+\alpha_Z}}, 
d(p)=\frac{B}{(p^2)^\kappa}, 
f(p)=\frac{C}{(p^2)^{\alpha_f}}\; ,
\ee
and $\kappa>0$ from the horizon condition. By means of the ghost DS equation given in Eq.\ (\ref{d})  one can relate the infrared exponent of the gluon to that of the ghost,
\be
\label{sumrule}
\alpha_Z+2\kappa=\frac{d-4}{2}
\ee
in $d+1$ dimensional Coulomb gauge YM theory. Thus, we eliminate $\alpha_Z$ in favor of $\kappa$. Using the infrared integral approximation, we derive from Eq.\ (\ref{d}) that \cite{SchLedRei06}
\be
\label{ghost2}
(p^2)^\kappa=(p^2)^\kappa AB^2N_cI_G(\kappa)
\ee
becomes exact for $p\rarr 0$, where
\be
\label{IG}
I_G(\kappa)=- \frac{4^\kappa (d-1)}{(4\pi)^{d/2+1/2}} \,\frac{\Gamma(\frac{d}{2} - \kappa )\,\Gamma(-\kappa )\,
      \Gamma(\frac{1}{2} + \kappa )}{\Gamma(\frac{d}{2} - 2\,\kappa )\,\Gamma(1 + \frac{d}{2} + \kappa )}.
\ee
Furthermore, we can infer from Eq.\ (\ref{ghost2}) that 
\be
\label{ABG}
AB^2N_cI_G(\kappa)=1
\ee
has to hold. Simultaneously solving the gluon DS equation with $\omega$ dominated by the curvature $\chi(p)$ in the infrared \cite{Feuchter:2004mk} we find for $d=3$ exactly two solutions that comply with the horizon condition \cite{SchLedRei06,Zwa02},
\be
\label{resk}
\kappa\in\left\{0.398,\frac{1}{2}\right\} .
\ee
Both of these solutions have been found numerically in refs.
\cite{Feuchter:2004mk} and \cite{Epple:2006hv}, respectively.

We now aim at a solution for the Coulomb form factor DS equation given by Eq.\ (\ref{f}). For $p\rarr 0$, one finds
\be
\label{DSEf}
(p^2)^{-\alpha_f}=(p^2)^{-\alpha_f} AB^2N_cI_f(\kappa,\alpha_f)
\ee
where
\begin{widetext} 
\be
I_f(\kappa,\alpha_f)=\frac{(d-1)\kappa}{(4\pi)^{d/2}}\frac{\Gamma\left(\alpha_f\right)\Gamma\left(2\kappa\right)\Gamma\left(d/2-2\kappa-2\alpha_f\right)}{\Gamma\left(d/2-2\kappa\right)\Gamma\left(d/2-\alpha_f+1\right)\Gamma\left(\alpha_f+2\kappa+1\right)} .
\ee
\end{widetext}
From Eq.\ (\ref{DSEf}) we can see that 
\be
\label{ABf}
AB^2N_cI_f(\kappa,\alpha_f)=1
\ee
has to be fulfilled. The above relation from the Coulomb form factor
DS equation can now be plugged into the relation Eq.\ (\ref{ABG}) from
the ghost DS equation to find 
\be
\label{condition}
I_G(\kappa)=I_f(\kappa,\alpha_F) .
\ee

We now specify the spatial dimension $d$. For $d=3$, Eq.\ (\ref{condition}) yields
\begin{widetext} 
\be
\label{solf}
1= \frac{\left( -1 + 2\,\kappa  \right) \,\cos (\pi \,\left( \alpha_f  + 2\,\kappa  \right) )\,\Gamma(4 - 2\,\alpha_f )\,\Gamma(-2\,\kappa )\,
    \Gamma(1 + 2\,\alpha_f  + 4\,\kappa )\,\sin (\pi \,\alpha_f )}{\pi \,\left( -1 + \alpha_f  \right) \,\left( 3 + 2\,\kappa  \right) \,
    \left( -1 + 2\,\alpha_f  + 4\,\kappa  \right) \,\Gamma(2 + 2\,\kappa )} .
\ee
\end{widetext} 
The numerical solution to this equation gives $\kappa$ as a function of 
$\alpha_f$, as shown in Fig.\ \ref{kvonf}.
One can see immediately that for any value of $\alpha_f$, the ghost exponent $\kappa$ yields
\be
\label{restrict}
\kappa<\frac{1}{4}.
\ee
 Therefore, recalling the result Eq.\ (\ref{resk}), there exists no value
 of $\alpha_f$ for which all three DS equations are satisfied. 

It is instructive to focus on the case where 
\be
\alpha_f+2\kappa=1
\ee
since this leads directly to a linearly rising potential for static quarks. Plugging this constraint into 
Eq.\ (\ref{solf}), we find
\be
1=\frac{-2\,\kappa \,\left( 3 + 2\,\kappa  \right) \,\cos (2\,\pi \,\kappa )\,\Gamma(-1 - 4\,\kappa )\,\Gamma(1 + 2\,\kappa )}  {\left( -1 + 2\,\kappa  \right) \,\Gamma(-1 - 2\,\kappa )}
\ee
which has the numerical solution $\kappa=0.245227$. Surprisingly, this result is exactly agreed upon 
by lattice calculations \cite{LanMoy04}, where $\kappa=0.245(5)$ was found.
However, one should notice that the lattice calculations carried out so far in
Coulomb gauge and also in Landau gauge use too small lattice to give reliable
results in the infrared.

\begin{figure}[h]
\includegraphics{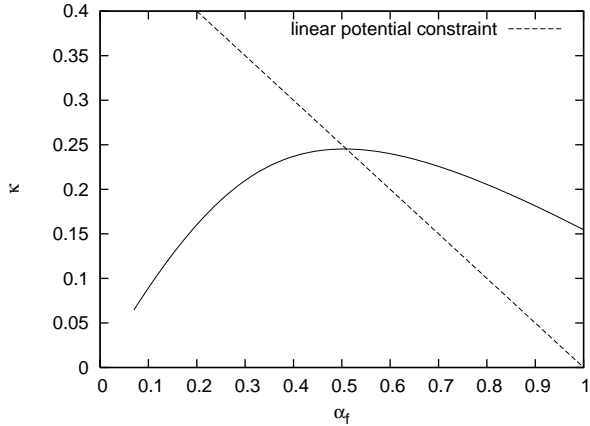} 
\caption{The ghost exponent $\kappa$ as a function of $\alpha_f$ is shown by a solid
line for $3+1$ dimensions. As a dashed line, it is indicated for which values a linear potential will
arise.} 
\label{kvonf}
\end{figure}

In $2+1$ dimensions, the calculations are equivalent. From a
simultaneous treatment of the ghost DS equation and the gluon DS
equation we find for the assumptions that the curvature dominates the
infrared a unique solution for the ghost exponent \cite{Zwa02},
\be
\kappa=\frac{1}{5} .
\ee
Note that with angular approximation the result is $\kappa=1/4$. In a
recent publication \cite{Feuchter:2007mq}, this value for $\kappa$ was confirmed numerically. On the other hand, if 
we consider only the ghost DS equation and the equation for the Coulomb form factor, 
see Eqs.\ (\ref{d}) and (\ref{f}) for $d=2$, one gets by 
enforcement of the condition Eq.\ (\ref{condition}):
\be
\label{fd2}
\frac{\Gamma(\alpha_f )\,\Gamma(1 - \alpha_f  - 2\,\kappa )\,\Gamma(\kappa )\,\Gamma(2 + \kappa )}
  {\Gamma(2 - \alpha_f )\,{\Gamma(-\kappa )}^2\,\Gamma(1 + \alpha_f  + 2\,\kappa )}=1\,  .
\ee
The numerical solution is shown in Fig.\ \ref{kvonf2}. If we require the potential to be linearly rising, the solution has to obey
$\alpha_f+2\kappa=\frac{1}{2}$.
Eq.\ (\ref{fd2}) then leads to the numerical value of
$\kappa=0.138$. Lattice results do not agree with this value \cite{Moy}, although they do for $3+1$ dimensions.

\begin{figure}[h]
\includegraphics{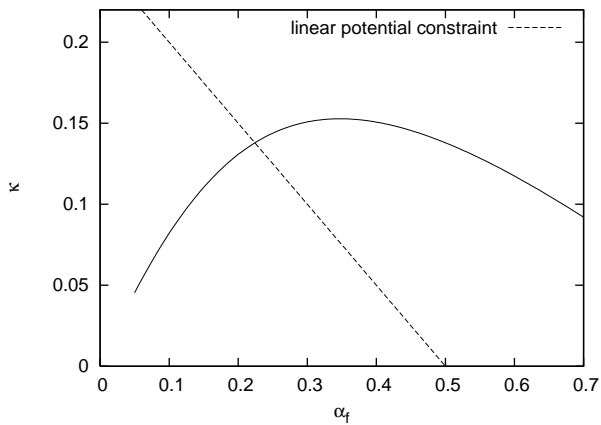}  
\caption{Same as Fig.\ \ref{kvonf}, here in $2+1$ dimensions.} 
\label{kvonf2}
\end{figure}

It is interesting to note that, leaving the DSE for $\omega$ aside, a
restriction on the infared exponents (\ref{restrict}) arises
from self-consistency of the equations for $d$ and $f$ only. This
restriction also concerns the infrared behaviour of $\omega$ via the
sum rule (\ref{sumrule}). Consider a fixed wave functional with a
kernel $\omega$ that yields $\kappa>\frac{1}{4}$ by its infrared behaviour. In
principle, it should be possible to calculate $d$ and $f$ for such a
wave functional. However, due to the restriction (\ref{restrict}), no solution can
be found. We therefore conclude that the Eq.\ (\ref{restrict}) must be due to
the approximations made in the DSEs for $d$ and $f$. Nevertheless, let us
stick to the approximation scheme here and investigate whether the
relaxation of the horizon condition allows for a self-consistent
solution of all DSEs.

\section{Full subcritical solutions}
\label{Numerical} 
For the IR analysis given above it follows that the Dyson-Schwinger equations, emerging from the minimization of the vacuum energy functional do not admit coupled solutions with all the functions being IR enhanced. Therefore in the following we do not implement the horizon condition, but rather chose an infrared finite ghost propagator. This allows us to include the ghost propagator fully in the equation for the Coulomb form factor.

Technically, the system is solved in the same way as described in \cite{Epple:2006hv, Reinhardt:2007wh}, except for the following differences. First, the finite renormalization constant $d^{-1}_0$ is implemented. Then, the infrared extrapolations of the form factors no longer can be done using power law ans\"atze, since the form factors are all infrared finite. Instead, we approximate the form factors using ans\"atze of the form
\begin{align}
\label{ir-ansatz-d}
d_\mathrm{ext}(k\to 0) = \frac{1}{k^\beta/B+b}
\end{align}
where $\beta$, $B$ and $b$ are parameters and extracted using least-squares fitting. One sees immediately that this ansatz corresponds to the previous ansatz $Bk^{-\beta}$, see (\ref{ir-ansatz-schl-d}), if one sets $b=0$, and that with a finite $b$ we get for the infrared limit of this extrapolating ansatz
\begin{align}
\lim_{k\to 0} d_\mathrm{ext}(k\to 0) = \frac{1}{b},
\end{align}
which means that the fit parameter $b$ corresponds to the renormalization parameter $d^{- 1}_0$. It can serve as a check that the resulting fit parameter $b$ actually turns out to be (numerically) equal to the (input) renormalization parameter $d^{- 1}_0$.

If not stated otherwise below, the other renormalization parameters are chosen
as $\mu=1$,
$f_\mu=1$. Physical units are not available yet in a strict sense, since the usual method of
adjusting the Coulomb string tension to its physical value can not be employed here, since the resulting quark-antiquark potential is not linearly rising. This is a direct consequence of the finite renormalization constant $d^{-1}_0$.

Ans\"atze of the same form like (\ref{ir-ansatz-d}) are then employed for the infrared extrapolations of the
other form factors $\chi(k)$, $f(k)$, $\omega(k)$. We chose to use the
same ans\"atze here because the numerically obtained values of the form factors can
be approximated by this ansatz pretty well. Of course, every form factor has its own set of fitting parameters. Also $\chi(k)$ and $\omega(k)$ are fitted separately (instead of using a single fit with the combined data, as it has been successful in solving the system with the horizon condition).

\subsection{The search for the critical $d(0)$}
First, we searched for the maximal value of $d(0)$ for which the system still
has a solution. If one could carry this value to infinity, one had again the
horizon condition. It is known from the analytical calculations that the system,
however, has no consistent solution, once the horizon condition is enforced.
Thus, we expect the system to have a solution only for values of $d(0)$ smaller
than some critical value.

Results are presented in Figs.\ref{fig-res-ghost}-\ref{fig-res-with-horizon-omchi}. In 
Fig.\ \ref{fig-res-ghost}, we show the ghost form factor $d(k)$.
One finds that the numerical solutions actually show the desired behavior in the
infrared, that the infrared fixpoint is at the value of the renormalization
constant $d^{- 1}_0$. In Fig.\ \ref{fig-res-coulomb}, we show the Coulomb form
factor $f(k)$. One finds that with larger and larger values of  $d_0$, the
Coulomb form factor $f(k)$ becomes more and more 
enhanced. This has some limit, though, as it is not possible to obtain a
solution for arbitrary small values of $d^{- 1}_0$. Furthermore, as the Coulomb
form factor gets more and more enhanced, it  becomes negative for large momenta. 
This behavior is not desireable. This fact does not seem to depend sensitively on $f(\mu)$. The largest $d(0)$ where both $f(k)$ is positive for all $k$ and the solution for $d(k)$ is stable has been found to be $d^{-1}_0 \approx 0.02$.

\begin{figure}[h]
\includegraphics{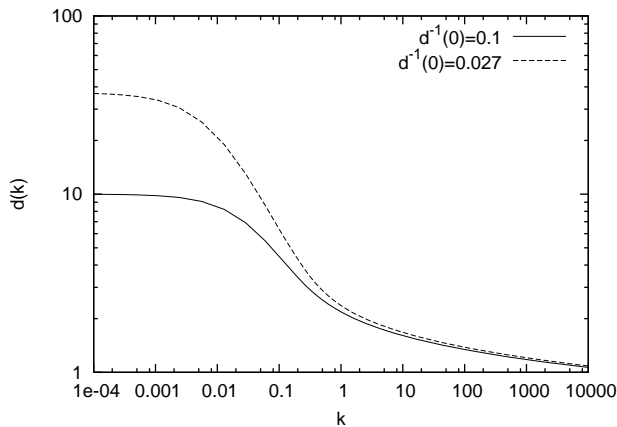} 
\caption{The ghost form factor $d(k)$.\label{fig-res-ghost} } 
\end{figure}
\begin{figure}[h]
\includegraphics{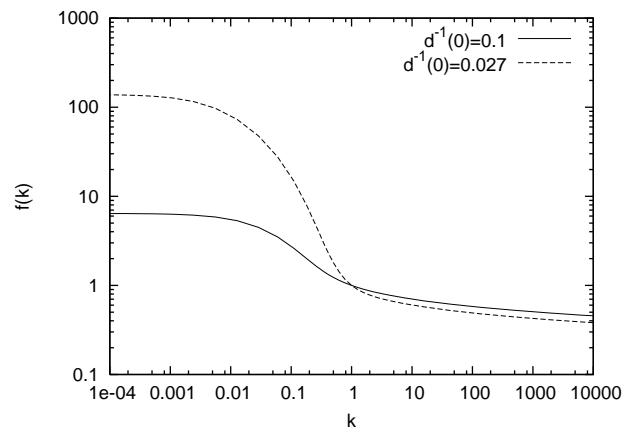} 
\caption{The Coulomb form factor $f(k)$.\label{fig-res-coulomb} } 
\end{figure}

In Fig.\ \ref{fig-res-omega-chi} we show the results for $\omega(k)$ and $\chi(k)$. We find that these are infrared finite (perhaps with logarithmical corrections), and not diverging like a power law, as it was found when using the horizon condition.

\begin{figure}[h]
\includegraphics{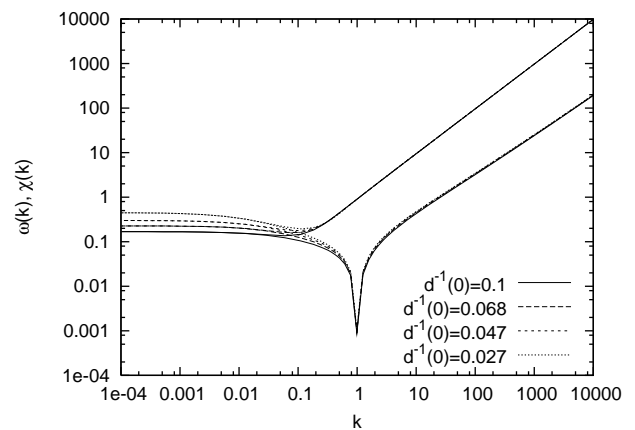}
\caption{The gluon energy $\omega(k)$ and the modulus of the
scalar curvature, $|\chi(k)|$. The scalar curvature $\chi(k)$ is positive in the
infrared, negative in the ultraviolet, such that in the log-log-plot, it seems
like it has a pole at its zero.\label{fig-res-omega-chi} } 
\end{figure}

\begin{figure}[h]
\includegraphics{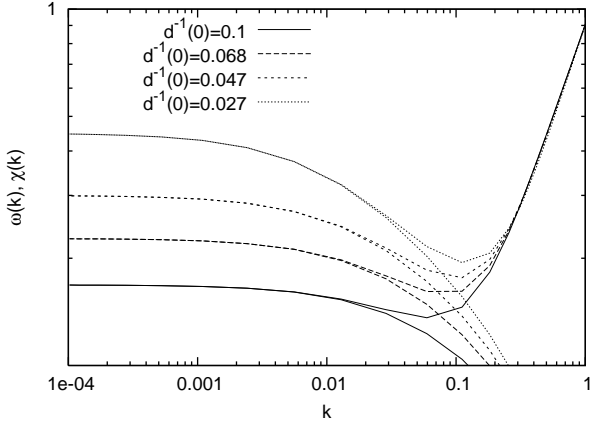}
\caption{Infrared part of the plot shown in Fig.\ \ref{fig-res-omega-chi}.\label{fig-res-omega-chi-IR} } 
\end{figure}

We have stressed so far the new infrared behavior of the solutions. As 
shown in Figs.\ \ref{fig-res-with-horizon-d} and \ref{fig-res-with-horizon-omchi}, the ultraviolet behaviour is the 
same as in the case with the 
horizon condition implemented. (However, these results are hard to compare, since the results with the horizon condition were obtained using different equations, which use the bare ghost propagator in the Coulomb equation, see above.)

\begin{figure}[h]
\includegraphics{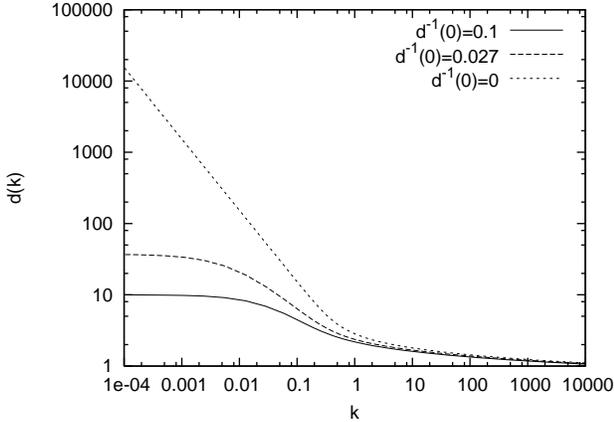} 
\caption{The ghost form factor $d(k)$, computed as described in the text, together with the
result described previously, where the horizon condition is implemented. Note
that the Coulomb form factor in the solutions with the horizon condition is
computed with the bare value of the ghost form factor, $d\equiv 1$, and thus is
not enhanced in the infrared.\label{fig-res-with-horizon-d} } 
\end{figure}

\begin{figure}[h]
\includegraphics{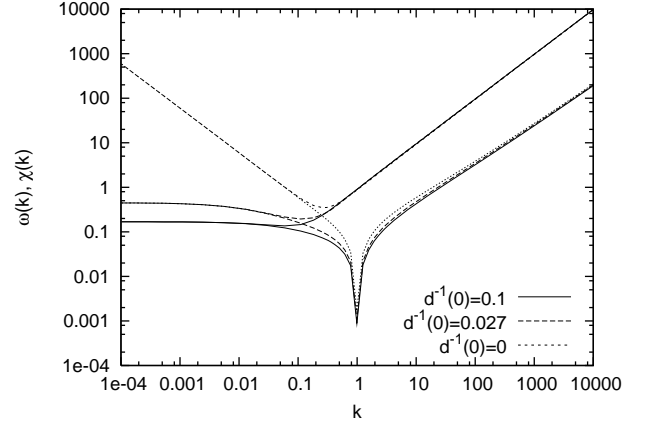}
\caption{C.f.\ Fig.\ \ref{fig-res-with-horizon-d}, but here shown are the gluon form factor, $\omega(k)$, and the modulus of the scalar curvature, $\chi(k)$.\label{fig-res-with-horizon-omchi} } 
\end{figure}

Furthermore, we take a look at the product $d^2(k)f(k)$, see Fig.\ \ref{fig-res-d2f}, which is related
 to the static quark potential in momentun space $V(k)$,
\begin{align}
d^2(k)f(k) \equiv V(k)k^2\:.
\end{align}
Thus, $d^2(k)f(k) \sim 1/k^2$ corresponds to $V(k) \sim 1/k^4$, which after
 Fourier transform yields a linearly rising quark potential. What one finds 
 in the numerical results is, of course, 
 that the static quark potential saturates for large distances, which stems from the fact that both $d(k)$ and $f(k)$ are infrared finite, but interesting is the feature that for a large momentum range, $d^2(k)f(k)$ actually roughly behaves like $1/k^2$, which results in a linear potential for a large range of distances. We may then assume the slope of this linear part as Coulomb string tension and introduce a physical scale this way. In Fig.\ \ref{fig-res-quarkpot} the result of $V(k)$ in physical units that were introduced this way is shown.

\begin{figure}[h]
\includegraphics{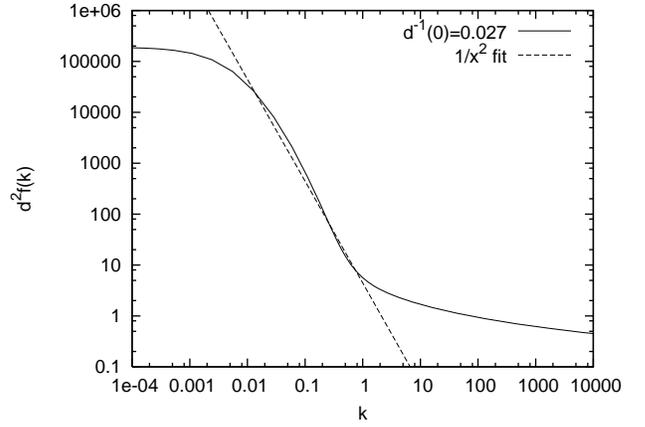} 
\caption{The product  $d^2(k)f(k)$.\label{fig-res-d2f} } 
\end{figure}

\begin{figure}[h]
\includegraphics{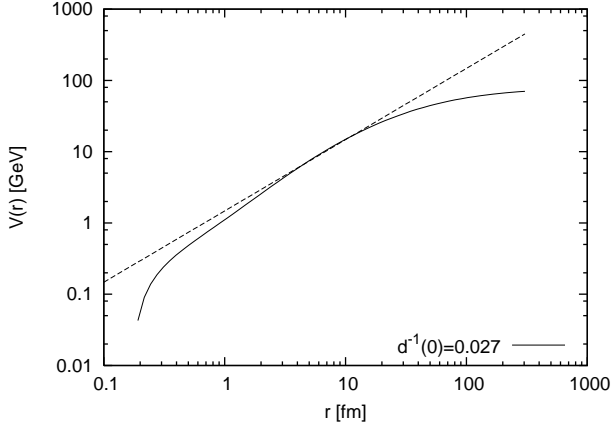}
\caption{The static quark potential in coordinate space, $V(r)$.\label{fig-res-quarkpot} } 
\end{figure}

\subsection{The search for the critical $f(\mu)$}

In a second calculation, $f(\mu)$ was increased to see whether a
solution where $f(k)$ becomes infrared divergent (''critical
solution'') can be found. However, increasing $f(\mu)$ over some
specific value produces a system where $\omega(k)$ becomes instable in
the infrared. This corresponds to the idea that a critical behaviour
leads to the non-availability of a fully consistent solution of the
three coupled equations. One can then use an additional degree of
freedom that arises when one does not set the finite constants on the
right hand sides of the Eqs.\ (\ref{24}) and (\ref{25}) to  zero, but
instead to some finite values. We rewrite the gap equation as
\begin{eqnarray}
\label{re11}
\omega^2 (k) - \bar{\chi}^2 (k)  & = & 
k^2 + \xi_0^\prime + \Delta I^{(2)}_\omega (k,0) \nonumber \\
& + & 2 \bar{\chi} (k) \left( \xi + \Delta I^{(1)}_\omega(k,0) \right) , 
\end{eqnarray}
where $\xi$ and $\xi^\prime_0$ are finite parameters. Setting these parameters to  zero  results in the same gap equation (\ref{30}) as used before. $\xi^\prime_0$ can be used to adjust the value of $\omega(\mu)$. In order to keep the (for the 't~Hooft loop desired) IR behavior of $\lim_{k\to 0} (\omega(k)-\bar\chi(k)) = 0$, $\xi$ must then assume a certain value,
\begin{align}
\label{re13}
 \xi = -\frac{\xi_0^\prime}{2 \bar{\chi} (0)}    \, .
\end{align}
Using this additional degree of freedom of tuning $\xi^\prime_0$, one can effectively increase $\omega(\mu)$, and also the infrared limit $\omega(0)$. 
This has the effect that the system becomes stable again. However, in the 
search of the critical $f(\mu)$, this does not help, since the infrared limit 
of $f(k)$ is decreased by the increased infrared 
strength of $\omega(k)$.

\begin{figure}[h]
\includegraphics{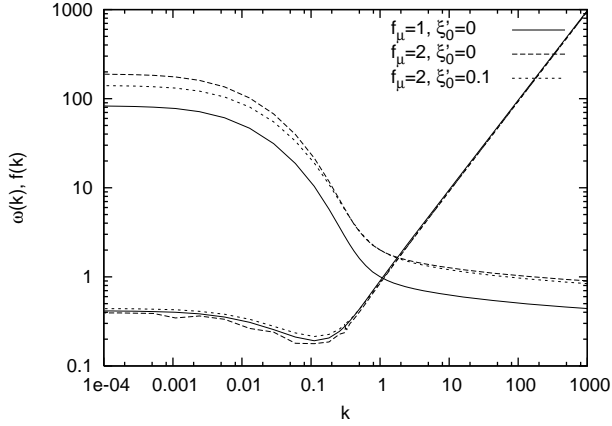}  
\caption{Solutions for the fully coupled system for different $f_\mu$ and
$\xi_0^\prime$.\label{fig-res-xi} } 
\end{figure}

\begin{figure}[h]
\includegraphics{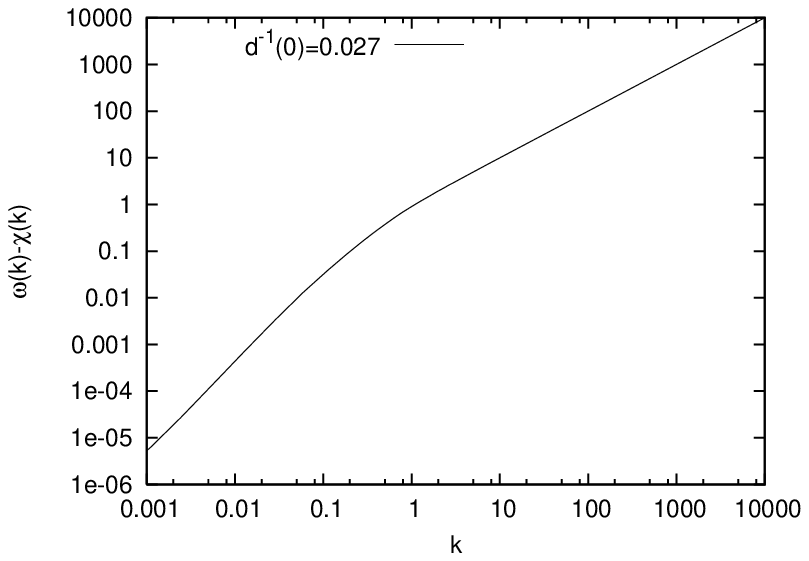}
\caption{$\omega(k)-\chi(k)$ log-over-lin plot.\label{fig-res-omega-minus-chi} } 
\end{figure}

\begin{figure}[h]
\includegraphics{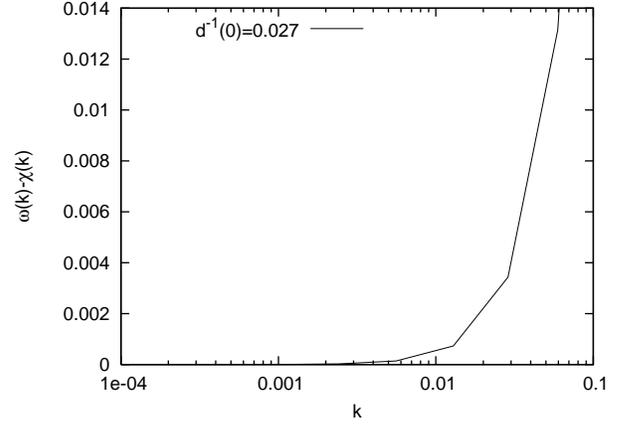}
\caption{$\omega(k)-\chi(k)$, lin-over-log plot.\label{fig-res-omega-minus-chi-ylin} } 
\end{figure}

\begin{figure}[h]
\includegraphics{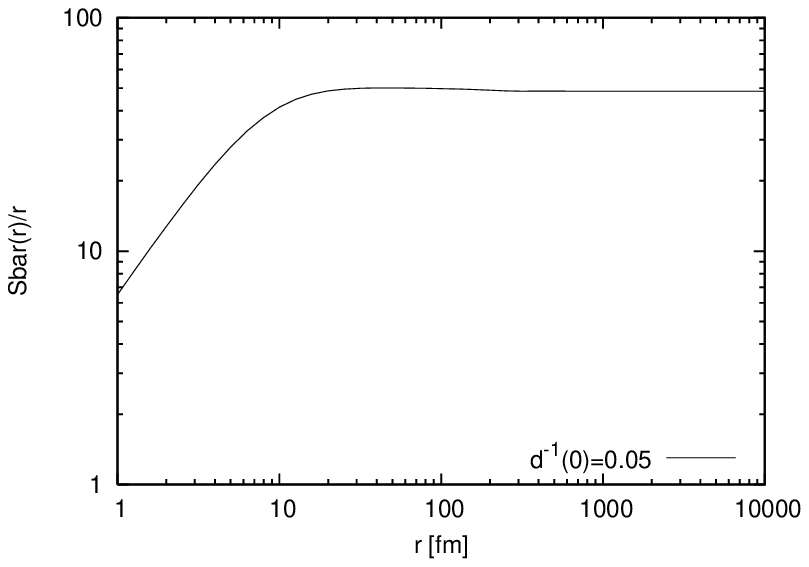}
\caption{The renormalized 't~Hooft loop exponent
$\bar{S}(R)/R$.\label{fig-res-thooft}  } 
\end{figure}

Fig.~\ref{fig-res-xi} illustrates this. We see the solution corresponding to 
Eq.\ (\ref{30}), with $f_\mu=1$. Increasing $f_\mu$ to $2$ produces
 the long dashed line, which shows instabilities in the infrared behaviour of 
 $\omega(k)$. Increasing $\xi_0^\prime$ as well, see the short-dashed line, we get stable solutions again, but with a smaller (non-critical) value of $f(k)$.

In another calculation, we test the point claimed in the introduction that the Coulomb equation is responsible for the non-availability of solutions of the fully coupled system when the horizon condition is included. To do so, we calculate some initial, infrared-finite $\omega(k)$ as solution of the subcritical system as shown in Fig.\ \ref{fig-res-omega-chi}. We then keep this $\omega(k)$ fixed and solve the system only for $d(k)$ with decreasing $d^{-1}_0$, and with these $d(k)$ we calculate the corresponding $f(k)$. We find that for every finite $d^{-1}_0$, both $d(k)$ and $f(k)$ have solutions. With decreasing $d^{-1}_0$, the $d(k)$ approach a certain asymptotical function. In the critical case where we perform the calculation with the horizon condition, $d^{-1}_0=0$, the ghost equation still has a solution, and this solution is equal to the limit of the solutions with finite, but decreasing $d^{-1}_0$,
\begin{align}
\lim_{d^{-1}_0\to 0} d(k; d^{-1}_0) =  d(k; d^{-1}_0=0)\, ,\label{asymp-d-k}\\
\intertext{with}
d(k; 0) < \infty\quad \forall\, k>0.
\end{align}
On the other hand, the $f(k)$ become larger and larger with decreasing $d^{-1}_0$, and an asymptotical function $f(k; d^{-1}_0=0)$ that is defined similar to (\ref{asymp-d-k}) does not exist. In the critical case where the horizon condition is implemented, the Coulomb equation has no solution: during the iteration process, the values for $f(k)$ grow larger and larger, without limit, to diverge eventually, representing the fact that the Coulomb equation has no solution in this case. Results of this computations are shown in Figs.\ \ref{fig-no-horizon-fixed-omega-d},\ref{fig-no-horizon-fixed-omega-f}.

\begin{figure}[h]
\includegraphics{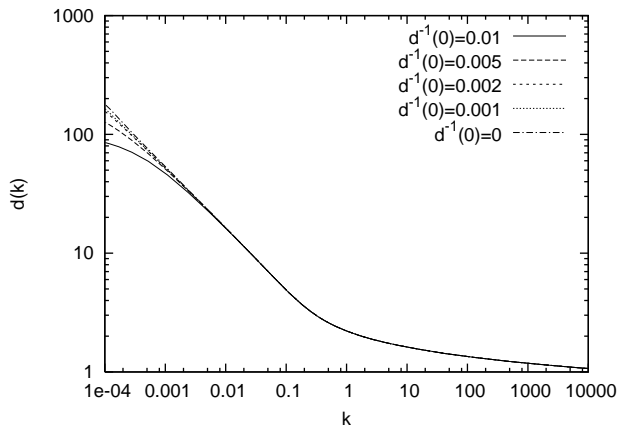}
\caption{The ghost form factor for different values of $d^{-1}_0$ calculated with a fixed, infrared-finite $\omega(k)$. We find that this form factor approaches a limit that corresponds to the critical solution with implemented horizon condition.\label{fig-no-horizon-fixed-omega-d}}
\end{figure}

\begin{figure}[h]
\includegraphics{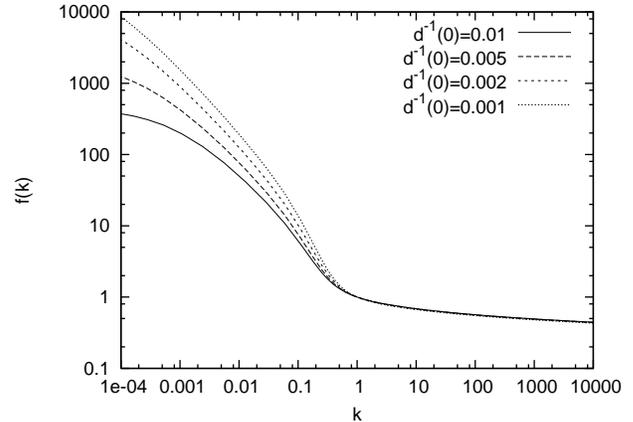}
\caption{The Coulomb form factors, calculated with a fixed, infrared-finite $\omega(k)$, and the ghost form factors as shown in Fig.\ \ref{fig-no-horizon-fixed-omega-d}. This form factor does not have a limit.\label{fig-no-horizon-fixed-omega-f}}
\end{figure}

Finally, we calculated the 't~Hooft loop using the solutions for the form factors presented in this paper.
It has been shown in \cite{Reinhardt:2007wh} that the large-$R$-behavior of the 't~Hooft loop is determined by the infrared behavior of $\omega(k)-\chi(k)$. We give results for this quantity $\omega(k)-\chi(k)$, and for the 't~Hooft loop itself, in Figs.\ \ref{fig-res-omega-minus-chi}-\ref{fig-res-thooft}. If we choose our renormalization constants such that $\omega(k)-\chi(k)$ approaches $0$ for $k\to 0$, then the 't~Hooft loop goes to a constant for large $R$, which is in accord with the results given in \cite{Reinhardt:2007wh}.

\section{Summary}
\label{Summary}
A non-perturbative solution to the Yang-Mills Schr\"odinger equation
in Coulomb gauge was presented. As a variational ansatz, a Gaussian
type of wave functional was used to minimize the energy. The
ultraviolet divergences in the three Dyson-Schwinger equations thus
obtained were removed by introduction of appropriate counter terms in
the Hamiltonian and in the wave functional. 
It was shown that within the approximations, no
self-consitent solution to all Dyson-Schwinger equations can be found
that complies with the horizon condition. Relaxing the horizon
condition, on the other hand, we did find a self-consistent solution
where all form factors approach finite values in the infrared. This
subcritical solution arises if the gauge coupling is chosen to be
smaller than some critical value. Approaching this critical value from
below, the Coulomb form factor $f$ ceases to solve its DSE. Apparantly, the approximation of the
ghost-gluon vertex has a different effect on $f$ than it has on
$d$. This explains why in previous calculations, critical solutions
were found only if the DSE for $f$ was ignored.

The subcritical solution with infrared finite form factors presented
here have some interesting features. First of all, the gluon
propagates like massive particle. Secondly, with the ghost and Coulomb
form factors $d$ and $f$ being infrared finite, the heavy quark
potential can be found to be linear just for a finite range of
separation, up to $10fm$. For phenomenological calculations, the use of
these form factors may give sensible results, and a further
investigation would certainly be interesting. 

\section*{Acknowledgements}
This work was supported by the Deutsche Forschungsgemeinschaft (DFG)
under contract no.\ Re856/6-1,2 and the US Department of Energy
under grant DE-FG0287ER40365. The authors are grateful to Axel Weber for valuable
discussions. APS would also like to acknowledge the hospitality extend during his visit  to T\"ubingen University where part of this work was completed.


\begin{thebibliography}{99}

 \bibitem{Thompson:1997bs}
  D.~R.~Thompson {\it et al.}  [E852 Collaboration],
  Phys.\ Rev.\ Lett.\  {\bf 79}, 1630 (1997)
  [arXiv:hep-ex/9705011].


  \bibitem{Ivanov:2001rv}
  E.~I.~Ivanov {\it et al.}  [E852 Collaboration],
  Phys.\ Rev.\ Lett.\  {\bf 86}, 3977 (2001)
  [arXiv:hep-ex/0101058].

  \bibitem{Szczepaniak:2003vg}
  A.~P.~Szczepaniak, M.~Swat, A.~R.~Dzierba and S.~Teige,
  Phys.\ Rev.\ Lett.\  {\bf 91}, 092002 (2003)
  [arXiv:hep-ph/0304095].

  \bibitem{Adams:1998ff}
  G.~S.~Adams {\it et al.}  [E852 Collaboration],
  Phys.\ Rev.\ Lett.\  {\bf 81}, 5760 (1998).

\bibitem{Christ:1980ku}
  N.~H.~Christ and T.~D.~Lee,
  Phys.\ Rev.\  D {\bf 22}, 939 (1980)
  [Phys.\ Scripta {\bf 23}, 970 (1981)].
  
  \bibitem{Swift:1988za}
  A.~R.~Swift,
  Phys.\ Rev.\  D {\bf 38} (1988) 668.
  
  
  \bibitem{Schutte:1985sd}
  D.~Schutte,
  Phys.\ Rev.\  D {\bf 31} (1985) 810.
  
  \bibitem{Cutkosky:1987yi}
  R.~E.~Cutkosky and K.~C.~Wang,
  Phys.\ Rev.\  D {\bf 37}, 3024 (1988).
  
  \bibitem{Zwanziger:2003cf}
  D.~Zwanziger,
  Phys.\ Rev.\  D {\bf 69}, 016002 (2004)
  [arXiv:hep-ph/0303028].
  
  
  \bibitem{Szczepaniak:2001rg}
  A.~P.~Szczepaniak and E.~S.~Swanson,
  Phys.\ Rev.\  D {\bf 65}, 025012 (2002)
  [arXiv:hep-ph/0107078].
  
  
  \bibitem{Szczepaniak:2003ve}
  A.~P.~Szczepaniak,
  Phys.\ Rev.\  D {\bf 69}, 074031 (2004)
  [arXiv:hep-ph/0306030].
  

\bibitem{Feuchter:2004mk}
  C.~Feuchter and H.~Reinhardt,
  Phys.\ Rev.\  D {\bf 70}, 105021 (2004)
  [arXiv:hep-th/0408236], [arXiv:hep-th/0402106].

\bibitem{Reinhardt:2004mm}
  H.~Reinhardt and C.~Feuchter,
  Phys.\ Rev.\  D {\bf 71}, 105002 (2005)
  [arXiv:hep-th/0408237].
  
 
  
  \bibitem{Greensite:1979yn}
  J.~P.~Greensite,
  Nucl.\ Phys.\  B {\bf 158}, 469 (1979).
  
  \bibitem{Greensite:2007ij}
  J.~Greensite and S.~Olejnik,
  arXiv:0707.2860 [hep-lat].
  
  \bibitem{Greensite:2004ke}
  J.~Greensite, S.~Olejnik and D.~Zwanziger,
  Phys.\ Rev.\  D {\bf 69}, 074506 (2004)
  [arXiv:hep-lat/0401003].
  
\bibitem{Greensite:2003xf}
  J.~Greensite and S.~Olejnik,
  Phys.\ Rev.\  D {\bf 67}, 094503 (2003)
  [arXiv:hep-lat/0302018].
  
  
  
\bibitem{Reinhardt:2007wh}
  H.~Reinhardt and D.~Epple,
  Phys.\ Rev.\  D {\bf 76} (2007) 065015
  [arXiv:0706.0175 [hep-th]].

\bibitem{SchLedRei06}
  W.~Schleifenbaum, M.~Leder and H.~Reinhardt,
  Phys.\ Rev.\ D {\bf 73} (2006) 125019
  [arXiv:hep-th/0605115].

\bibitem{Zwa02}
  D.~Zwanziger,
  Phys.\ Rev.\ D {\bf 65} (2002) 094039
  [arXiv:hep-th/0109224].



\bibitem{Epple:2006hv}
  D.~Epple, H.~Reinhardt and W.~Schleifenbaum,
  Phys.\ Rev.\  D {\bf 75} (2007) 045011
  [arXiv:hep-th/0612241].

\bibitem{LanMoy04}
  K.~Langfeld and L.~Moyaerts,
  Phys.\ Rev.\ D {\bf 70} (2004) 074507
  [arXiv:hep-lat/0406024].

\bibitem{Feuchter:2007mq}
  C.~Feuchter and H.~Reinhardt,
  arXiv:0711.2452 [hep-th].

\bibitem{Moy}
  L.~Moyaerts, PhD thesis, Univ.\ of T\"ubingen (2004).

\end{thebibliography}
\end{document}